# Day Ahead Price Forecasting Models in Thin Electricity Market

Sayani Gupta
Business Intelligence Unit
PAYBACK (American Express Subsidiary)
Gurugram, India
gupta.sayani@gmail.com

Puneet Chitkara
Infrastructure Government and Healthcare (IGH)
KPMG Advisory Services Private Limited
Gurugram, India
puneet.chitkara@gmail.com

*Abstract*— Day Ahead Electricity Markets (DAMs) in India are thin but growing. Consistent price forecasts are important for their utilization in portfolio optimization models. Univariate or multivariate models with standard exogenous variables such as special day effects etc. are not always useful. Drivers of demand and supply include weather variations over large geographic areas, outages of power system elements and sudden changes in contracts which lead the players to access power exchanges. These need to be considered in forecasting models. Such models are observed to considerably reduce forecasting errors by outperforming other models under conditions, which are neither infrequent nor recur at defined intervals. This paper develops models for India and tests the utility of these models using Model Confidence Set (MCS) approach which picks up the "best" models. The approach has been developed for a power utility in India over a period of two years in live business environment.

## I. Introduction

Thin day ahead electricity markets (DAMs) are characterized by highly volatile prices and the complication in understanding price variations increases when infrastructure is fragile. For almost up to a year after commencement of the exchanges in India, there were many trading time blocks where buyers and sellers were not there. More players started participating in these markets within a year. DAMs commenced operation in August 2008 in India and the share of these markets has grown to 3% since then. Electricity is traded in 15-minute time blocks and prices are discovered through a double side blind, uniform market clearing price auction mechanism. Approximately 3500-4500 MW is traded in each block of time and average flow in the grid is 125 GW (approx.) with a peak of 160 GW (approx.). Over last two-three years, DAM has been a buyer's market with considerable seller's offers remaining unsold. Commercially it is only prudent for power utilities in India to purchase/sell power in the DAM – and fortunately, this is expected to reduce overall power purchase costs, align long term, medium term and DA markets much better, and is also politically desirable. Price forecasts, therefore, assume importance. Univariate models – ARIMA, Seasonal ARIMA, ARFIMA, ARMA GARCH [1-2] and machine learning based models (SVM, ANN etc.) [3], [4] and Wavelet-ANN/ANFIS-PSO [5] based models have been deployed but seem to work well when weather and infrastructure conditions are relatively stable. Shahidehpour has mentioned [6] that system outages, weather conditions etc. need to be taken into account because outage of even one generator of 500 MW – or associated transmission link - (especially in markets where infrastructure is fragile) can lead to dramatic changes in prices. This has been observed in the Indian context quite often and needed to be modelled to enable consistently accurate price forecasts.

## II. Background

The work presented in this paper has been developed in a live commercial environment. There are two power exchanges in India – Indian Energy Exchange (IEX) and Power Exchange of India Limited (PXIL). IEX enjoys 99% market share in DAM. This work draws upon IEX prices for its analysis. DAM price movements in India on IEX are illustrated in Fig. 1. These prices are a result of interaction of supply and demand - which is geographically highly dispersed and hence dependent on not only weather conditions over a large geographic area but also on the availability of infrastructure (for example, mines getting flooded and hence the impact on prices). Also, each distribution utility (DISCOM) that bids on the exchange has a large geographic spread – weather, therefore, cannot be a single vector of temperature, humidity, cloud cover, wind speed and wind direction. DISCOMs, who offer to supply on the exchange procure most of their power through long/medium/short term bilateral contracts. Some of the generators committed to supply under bilateral contracts may lose their generation to unpredictable outages. Further, contracts may end (especially short term – daily, monthly and up to three monthly) at dates which do not have a pattern. In order to serve power, DISCOMs may purchase in DAM (depending on their financial capability (or regulatory allowance) of arranging short term capital for the same) at irregular intervals in lieu of contracted power. Similarly, if a sale contract ends, traders, generators and the DISCOMs may suddenly offer on the exchange. Such uncertainties due to the dates of contractual agreements - which cause a change in supply/demand on the exchange by even 100 MW have a huge impact on prices. Since DISCOMs are not obliged to supply always, they may shed demand. Incidentally, there is no pattern. While a bottom-up engineering model of the power system – which tracks all transmission lines, nodal generation and demand and hence flows on the network may help, but the same requires considerable data – most of which is not in public domain in India. With this background, a statistical model, considering all the factors which explains price volatility, was conceived. The proposed model structure is explained in section-III.

## III. PROPOSED IDEA

The observed series in Fig.1 reflects the volatility in electricity prices. The Trigonometric Box-Cox ARMA Trend Seasonal [4] (TBATS(1, {1,5}(ARMA specifications of the residuals), 0.803(damping factor), {<96,5>, <672,4>, <35064,5>}) decomposition is shown in other five panels of Fig 1. The series is indicated to have a daily pattern with 5 significant Fourier terms (season1), weekly pattern (4 Fourier terms) and yearly pattern (5 Fourier Terms).

Forecasts based on the above decomposition results in a MAPE of between 12%-18% depending on alternative assumptions. This indicates that beyond the seasonal patterns identified above, there are other exogenous factors which drive prices.

The following are identified as the fundamental drivers of price on the Exchange: (i) 15-minute demand forecasts for major trading DISCOMs in India based on weather data (ii) Changes in Short/ Medium/ Long term bilateral contracts of all the DISCOMs and those of generators that have a merchant capacity for sale on the exchange, (iii) Availability / Outages of state sector/ central sector / private sector plants, (iv) Capacity offers for sale on power exchange from all Independent power producers, (v) Expected transmission corridor availability.

For (i), real time demand data is regularly captured from [7], [8]. Demand forecasts are based on not only special day / week day/end effect but also weather data. Depending on the size of the state, districts which have weather sensitive demand (agriculture and residential demand in India is sensitive to weather) in each state were identified.

Weather data for 51 districts in 11 major Indian states is regularly captured from [9] and is fed as input into the model. Short/Medium/Long term bilateral contract data is regularly taken from [10-13] for all the states in India. Similarly, (iii), (iv) and (v) data are taken from [14-27].

Based on the above data, multivariate Time Series and Machine Learning models have been developed along with univariate models. The models used in this paper are:

### A. Univariate Models

*1) ARFIMA:* Auto Regressive Fractional Integrated Moving Average (used for modelling long memory processes): named: ARFIMA1 (3.5 years training data) and ARFIMA2 (1.5 years training data),

*2)* Holt Winters Exponential Smoothing Method named: HW_1: (1 year training period) [28,29],

*3)* ARMA (Auto Regressive Moving Average) model with GARCH (Generalized Auto Regressive Conditional Heteroscedasticity) errors named: ag (3.5 years training data),

*4)* ANN: Artificial Neural Network named: pred_ANN_nods, pred_ANN_nods_15, pred_ANN_nods_30 (with 45, 15 and 30 days training period respectively) and lagged DAM prices up to order 3,

*5)* SVM: Support Vector Machines named pred_SVM_nods_15, pred_SVMl_nods_15 (with 15 days training period and lagged DAM prices up to order 3 with Kernel function as Radial and Linear respectively)

### B. Standard Multivariate Models

*1)* ANN models named: pred_ANN_ds, pred_ANN_ds_15, pred_ANN_ds_30 (with 45, 15 and 30 days training period respectively), lagged DAM prices up to order 3, last day's demand supply gap on IEX and block effect,

*2)* SVM named: pred_SVM_ds_7, pred_SVM_ds_15 (with 7 and 15 days training period respectively and Kernel function: Radial), lagged DAM prices up to order 3, last day's demand supply gap on IEX and block effect,

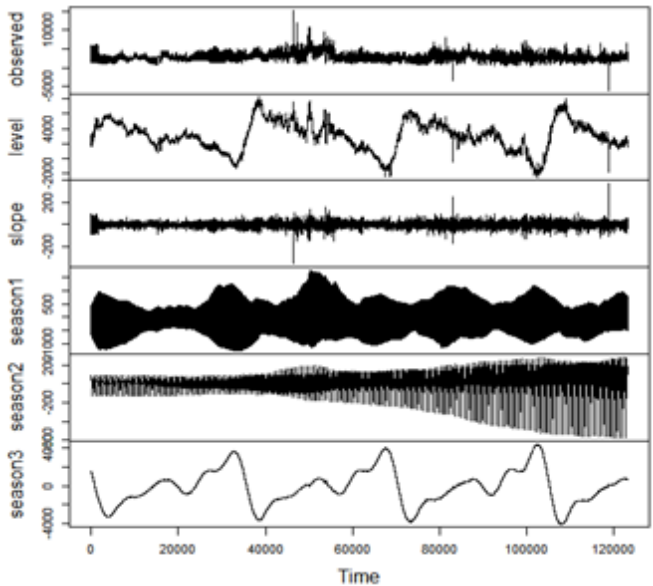

Fig. 1. Decomposition of price (15 minute block-wise series) for N3[1] from January 1, 2013 to September 25, 2016 into level, trend, daily (season1), weekly (Season 2) and yearly (season 3) seasonal components using TBATS (in R)

---

[1] Indian grid is divided in 12 congestion zones – Northern Region has three (N1, N2, N3), Eastern Region has two (E1, E2), North Eastern Region has two (A1, A2), Western region has three (W1, W2, W3), and Southern Region has two (S1, S2)

*3) SARIMAX: Seasonal Auto Regressive Integrated Moving average with last day's demand supply gap as the exogenous variable*

*C. Detailed Multivariate Models (contribution of this paper)*

The demand and supply variables for each utilities are taken into consideration in these category of models. Principal Component Analysis (PCA) has been employed to convert a set of possibly correlated variables into a set of values of linearly uncorrelated variables and identify the summarized factors that are critical to price formation among all the factors mentioned. The principal components which cumulatively explain 80% of the variation in the data are fed into the model. The factors which do not show any significant variation in the last 30 days are removed from the list of factors before conducting PCA as they won't play any role in impacting the prices.

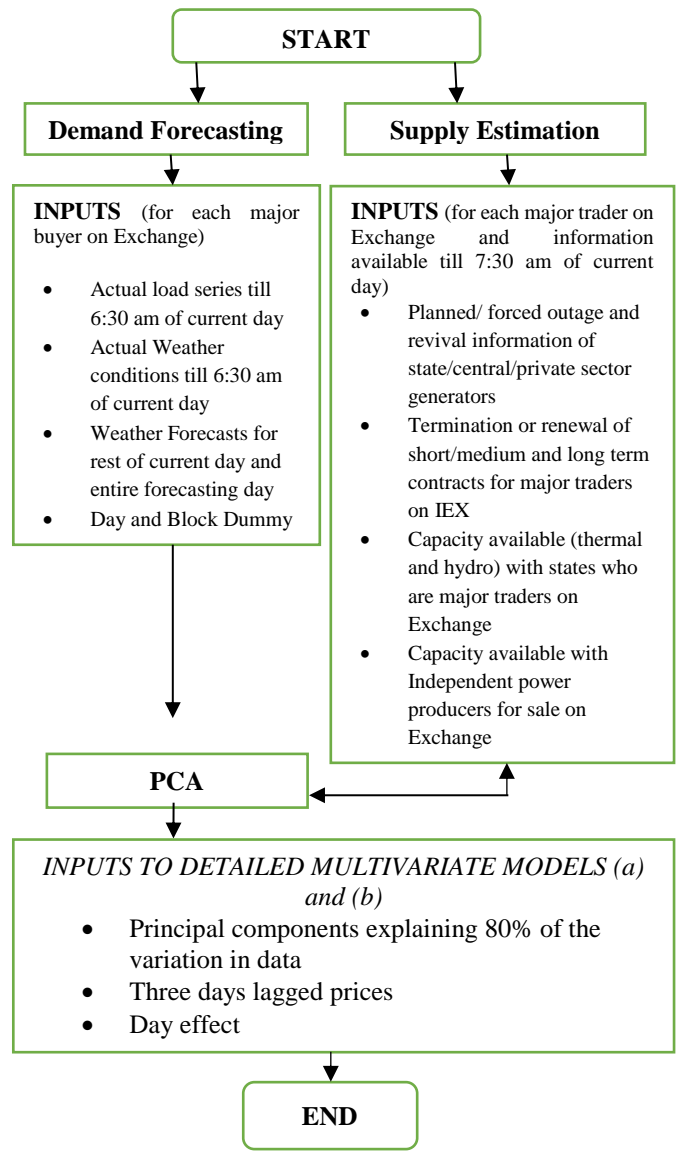

Fig. 2. Framework of Multivariate Model

The factors identified using PCA, are then fed into the following models:

- **Gradient Boosting Method (GBM) [30-31]** a machine learning technique that combines the strengths of two algorithms: regression trees (models that relate a response to their predictors by recursive binary splits) and boosting (an adaptive method for combining many simple models to give improved predictive performance). The models named: **Price Model** (number of trees5000), **Specf1** (number of trees: 6000) **and Specf1d** (along with last day's demand supply gap) are fed with factors from PCA day effect and three days lagged price series. Supply conditions of major IPPs have been additionally incorporated in the variable list considered above in the models named: **Price Model_ipp, Specf1_ipp**

- **SVM (Support Vector Machines):** SVM is a supervised machine learning algorithm which uses a technique called the kernel trick to transform the data and then based on these complex transformations it finds an optimal boundary between the possible outputs. Similar variables as in Specf1d named **svm_pca_15** and **svm_pca_30** (with Kernel: Linear and training period as 15 days and 30 days respectively). Additionally, supply conditions of major IPPs have been considered in the model **Svm_pca_15_ipp** and **Svm_pca_30_ipp**

Univariate models perform well under steady weather and system conditions, while detailed multi-variate models perform better than univariate models under conditions of varying weather and system availability (outage, change in contracts etc.). However, finally a model which provides consistent forecasts is helpful. There could be models which give very low Mean Absolute Percentage Error [6] in certain time blocks and on certain days but perform poorly on other days – such models are not practically usable, rather a model which results in "consistent" and "acceptable" level of MAPE are better suited in practical applications. Hence, the models are combined for each block of time using the concept of the Model Confidence Set (MCS) procedure developed in 2011 [33]. The procedure consists of a sequence of statistical tests which enable to construct a set of "best" models called "Superior Set Models" (SSM) with a given level of confidence. The null hypothesis of these tests assumes equal predictive ability (EPA) of all the models based on a loss function, which is, Mean Absolute Percentage Error, in this case. The Model Confidence Set procedure starts from an initial set of all forecasting models and results in a smaller set of superior models (i.e. SSM) based on their predictive ability over last 10 days. The predictions of these superior models are then combined with appropriate weights that equal the inverse of the loss value associated to each one of these models to give the final forecast of the day-ahead price.

Competing combination models are proposed in [34-36] However, MCS has an edge over other models as mentioned in [33].

## IV. Discussion

Consistency in model performance is assessed in terms of average daily MAPE, block-wise MAPE (Mean Absolute Percentage Error) and its variance. As mentioned earlier, univariate models perform well under "stable" conditions, however in thin markets detailed considerations, as discussed above, does improve predictive accuracy. We provide evidence that multivariate modelling approach provides an edge in predictive performance over univariate models across all datasets, seasons of the year or blocks of the day when the deviation in day-on-day prices during the same block of time is very high. As a measure of performance, usually average MAPE is reported over a day or week or month, low MAPEs are achieved with the exponential smoothing methods or other univariate methods, which may lead us to conclude that simpler and more robust methods, which require little domain knowledge, can outperform more complex alternatives. Such results are reported in [37, 38]. While, on an average daily/weekly/monthly basis the results of [37, 38] hold, one needs to analyze performance of these models during "abnormal" blocks of a day or some atypical days to appreciate the efficacy of multivariate models. Performance of univariate and multivariate models in Table I shows, for immediately past three months (November and December 2016 and January 2017) that when the variation on day-on-day prices (Lag_Diff in TABLE I) is low, that when the variation on day-on-day price in low, univariate models are more likely to be selected by the MCS procedure, whereas as this variation increases, the probability of multivariate models getting selected increases.

Table I is representative and shows day-on-day variations for E1. One might question the higher selection of univariate models when day-on-day block-wise difference in prices is in the 40%-60% range. Here, in the months of November and December, one observes weekly / ten-day trends – which are better captured by HW / other univariate models. For N3, it is observed that multivariate models perform better than univariate models in approximately 25% of the blocks on an average for these months which again establishes the importance of multivariate models over univariate models.

TABLE I. PERFORMANCE OF UNIVARIATE AND MULTIVARIATE MODELS FOR DIFFERENT RANGES OF DAY-ON-DAY PRICE VARIATION FOR E1 REGION

| Month | Models Chosen | Lag_Diff <20% | 20% < Lag_Diff < 40% | 40% < Lag_Diff <60% | Lag_Diff >60% |
|---|---|---|---|---|---|
| November | Univariate | 72.1% | 61.9% | 77.4% | 24.3% |
| | Multivariate | 27.9% | 38.1% | 22.6% | 75.7% |
| December | Univariate | 74.3% | 67.7% | 83.3% | 44.4% |
| | Multivariate | 25.7% | 32.3% | 16.7% | 55.6% |
| January | Univariate | 78.2% | 73.7% | 43.5% | 8.7% |
| | Multivariate | 21.8% | 26.3% | 56.5% | 91.3% |

To emphasize the utility of the multivariate models in the ensemble of models considered by the MCS model, we further demonstrate the same through examples presented in Fig. 2, Fig. 3, and Fig. 4 below. The examples have been picked up to demonstrate the efficacy of the results in different markets and different months.

Fig. 2 shows two days in which pattern of prices were unusually different from that of last day due to revivals/outages of plants, change in contracts or demand of utilities.

In Fig.2.demand in those blocks (15th Oct, 2015) fell abruptly compared to last day as drawl volumes from N2, W1 and S1 decreased considerably. Again, for 17th Sept, 2015, N2's (Rajasthan) drawl on power exchange decreased throughout the day as availability from central sector plants increased in the morning hours and demand was lower in the latter half of the day. Also, injection increased immensely from W2 (Maharashtra) due to lower demand and also higher availability owing to revival of state sector plants.

Univariate Models fails to anticipate such practical changes since forecasts are just based on previous price series. Fig.3. shows how multivariate models are a sharp improvement over univariate models for blocks in which changes over last day were huge. MAPE continued to remain high even for multivariate models in blocks where spikes were high, however, considerably less compared to univariate models. These are incidentally parts where even Fourier Transform or Wavelet based methods treat this variation as noise.

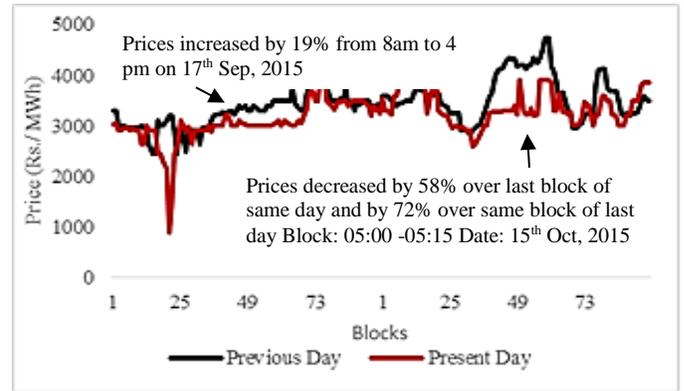

Fig. 3. Variation of prices on day ahead and block ahead basis. The first half shows prices for 14th and 15th Oct, 2015 (N3) and the next half shows price for 16th and 17th Sept, 2015(N3)

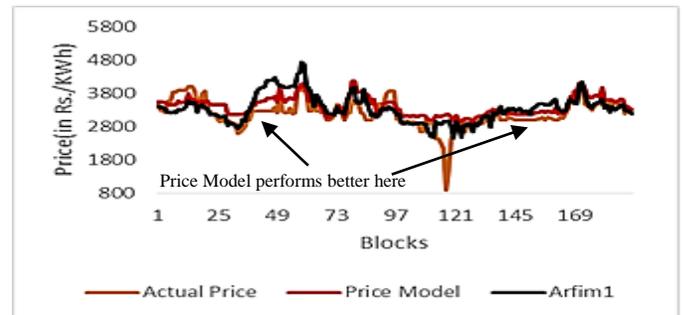

Fig. 4. Comparison of performance of multivariate models and univariate models for 17th September and 15th October, 2015 for N3 region

For example, Fig.4. shows similar comparison for E1 region. During the morning blocks (Block 1 to Block 20), the prices on 28-11-2016 were very low as compared to the prices on 29-11-2016.

During this period, smoothing Holts Winters model follows the prices on 28-11-2016 closely, while multi-variate model gives a better forecast.

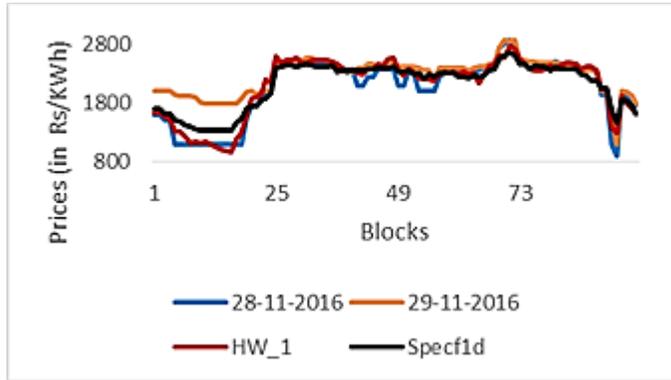

Fig. 5. Comparison of performance of multivariate models and univariate models for 28th and 29th November, 2016 for E1 region

Performance of Models across Blocks:

hrs., the combination model picks up detailed multivariate models (SVM_pca_15_ipp, Specf1d_ipp, Price Model_ipp) significantly during autumn and summer, and more so during summer months. Superiority of detailed multi-variate models during the day time and evening peak hours (especially between 18:00 – 22:00 hrs.) indicates the importance of detailed system based forecasting models developed in this work.

Consistency in terms of MAPE:

Finally, having demonstrated the importance of inclusion of multivariate models in our analysis and in the ensemble of models over which MCS operates, we demonstrate the efficacy of the overall process in terms of overall MAPE of the combination model.

These different competing models are built to answer specific econometric questions; hence their performance varies across blocks of the day and across seasons. Hence set of superior models in MCS varies with varying seasons and blocks. To illustrate the efficacy of the proposed technique, results corresponding to the four seasons of year 2015 / 2016 are presented. There are mainly five seasons in India, illustrative for a week in each season is shown in Table II.

| 00:00 -05:00 | 05:00 -10:00 | 10:00 – 15:00 | 15:00 – 18:00 | 18:00 – 24:00 |
|---|---|---|---|---|
| *Winter* - HW_1, ANN_ds_15, ANN_nods_15<br>*Late Autumn* - ANN_nods_15, SVM1_nods_15, SVM_nods_15<br>*Summer* - Specf1, Specf1d, SVM_ds_7, Ag | *Winter* -HW_1, SVM_nods_15, arfim1<br>*Late Autumn* - ANN_nods_30, arfim2, SVM_nods_15,arfim1<br>*Summer* - arfim1, HW_1, Price Model_ipp, ANN_nods_30 | *Winter* - SVM_nods_15, Price Model, arfim1<br>*Late Autumn* - ANN_nods, Price Model_ipp, SVM_nods_15, ANN_nods_15, ANN_nods_30<br>*Summer* - Price Model, arfim2, specf1_ipp, Specf1 | *Winter* - HW_1, ANN_nods_30, Price Model, Price, Model_ipp, Specf1 ,SVM_nods_15<br>*Late Autumn* - ANN_nods, ANN_nods_30, Price Model, Price Model_ipp, SVM_nods_15, Specf1<br>*Summer* - arfim1, HW_1, specf1_ipp, SVM_pca_15_ipp, SVM_pca_30, specf1d_ipp, Price Model, SVM_nods_15 | *Winter* - HW_1, Price Model_ipp, SVM_nods_15, specf1_ipp, ANN_nods_30 SVM_ds_15<br>*Late Autumn* - HW_1, SVM_ds_15, Ag<br>*Summer* - Price Model_ipp, specf1d_ipp, HW_1, Price Model, specf1_ipp, SVM_pca_30_ipp |

Fig. 6. Set of Superior Models in MCS across blocks of time and seasons.Winter, late autumn and summer week cover the period (1st December, 2015 -15th December, 2015), (16th October , 2015 -31st October, 2015) and (1st July, 2015 – 15th July, 2015) respectively

Fig. 6. shows that during 00:00 to 05:00 hrs. (when demand is relatively stable at the national/regional level and most of the demand is domestic, in winter and late autumn, electricity demand and supply conditions do not vary much) univariate models (arfim1, HW_1, ANN_nods_15) outperform other models. In summer and monsoon months (June end / July / August), fluctuations in weather conditions and system outages (leading to DISCOMs, Generators and Large Open Access consumers moving in and out of DAM) cause high variations in prices and hence detailed multivariate detailed models (Specf1, Specf1d) seen to perform better. From 05:00-10:00

TABLE II. PERFORMANCE OF COMBINATION MODEL ACROSS SEASONS AND REGIONS

| Season | Period | Region | MAPE |
|---|---|---|---|
| Fall | October 4, 2016 - October 10, 2016 | E1 | 4.7% |
| Winter | January 26, 2017- February 1, 2017 | E1 | 6.3% |
| Monsoon | Sept 13, 2016 - Sept 20, 2016 | E1 | 5.8% |
| Winter | January 28, 2015 - February 3, 2015 | N3 | 6.0% |
| Spring | April 3, 2015 - April 11, 2015 | N3 | 6.5% |
| Summer | June 11,2015 - June 17, 2015 | N3 | 11.8% |

## V. CONCLUSIONS

Power sector dynamics in terms of the outage of power system elements which serve bilateral contracts, changes in bilateral contracts and weather conditions over wide geographic regions impact electricity prices in DAMs. Variations in prices which would otherwise pass off as "noise" in most existing models (univariate and standard multivariate) are "captured" and accounted for by these models. The utility of these models is tested using Model Confidence Set (MCS) approach which picks up the "best" models for each block of time. Alternative specifications of such detailed multivariate models are shown to be frequently among the Superior Set of Models. One line of future work, where results could be further improved involves relating each component of the Wavelet Transform of the original price series to the variable considered in detailed Multivariate model and then using inverse wavelet transform to generate price forecasts.